\documentclass[11pt]{article}
\usepackage{amsfonts}

\newcommand{\sect}[1]{\setcounter{equation}{0}\section{#1}}
\newcommand{\subsect}[1]{\subsection{#1}}

\def\be{\begin{equation}}
\def\ee{\end{equation}}
\def\bea{\begin{eqnarray}}
\def\eea{\end{eqnarray}}

\def\R{{\mathbb R}}
\def\1{\'{\i}}                           

\def\rangospace{{r}}

\def\gal{{\cal{G}}}
\def\poinc{{\cal{P}}}
\def\desitter{{\cal{S}}}
\def\newhooke{{\cal{N}}}
\def\euclid{{\cal{E}}}

\def\ppa{{\cal C}_1}

\def\ppb{{\cal C}_2}
\def\jjj{{\cal J}}

\def\parity{\mathsf P}
\def\timereversal{{\mathsf T}}

\def\k{\kappa}
\def\ext{m}
\def\AA{\alpha}
\def\ww#1{W_{#1}}

\def\>#1{{\bf #1}}                 
 
\def\producto{\!\cdot\!}                 
\def\numcas{m}   
\def\comm#1#2{[\,{#1},\,{#2}\,]}              

\begin{document}
  
\hfill  
 
\bigskip

{\large{\bf{(Anti)de Sitter/Poincar\'e symmetries and representations }}}
 
{\large{\bf{
from Poincar\'e/Galilei through a classical deformation approach}}}

 \bigskip\bigskip

\begin{center}
Francisco J. Herranz$^\dagger$ 
and Mariano Santander$^\ddagger$
\end{center}

\begin{center}
{\it $^\dagger$ Departamento de F\'{\i}sica, Escuela
Polit\'ecnica Superior\\ 
Universidad de Burgos, E--09001 Burgos, Spain}
\end{center}

\begin{center}
{\it $^{\ddagger}$ Departamento de F\'{\i}sica Te\'orica,
Facultad de Ciencias\\ Universidad de Valladolid,
E--47011 Valladolid, Spain}
\end{center}

\bigskip\bigskip

\begin{abstract}
A classical deformation procedure, based on universal
enveloping    algebras, Casimirs and curvatures of symmetrical homogeneous
spaces,  is applied to several cases of physical relevance. 
Starting from the
$(3+1)$D Galilei algebra, we describe at the level of representations  the process
leading to its two physically meaningful deformed neighbours. The Poincar\'e algebra is
obtained by  introducing a negative curvature in the flat  Galilean phase space (or
space   of worldlines), while keeping a flat spacetime. To be precise, starting from a
representation of the Galilei algebra with both Casimirs {\it different} from zero, we
obtain a representation of the Poincar\'e algebra with both Casimirs {\it necessarily}
equal to zero.  The  Poincar\'e angular momentum,   Pauli--Lubanski components, position and
velocity operators, etc.\ are expressed in terms of `Galilean' operators through some
expressions deforming the proper Galilean ones.  Similarly, the Newton--Hooke algebras
appear by  endowing  spacetime with a  non-zero  curvature, while keeping a flat phase
space.  The same   approach, starting from the $(3+1)$D Poincar\'e algebra provides
representations of the (anti)de Sitter as Poincar\'e deformations.
\end{abstract}

 \bigskip\medskip 



\bigskip\medskip 

\newpage 


\sect{Introduction}

Classical deformations of Lie algebras can be considered as the opposite
process to  Lie algebra contractions. In general, starting
from a Lie algebra a {\em contraction} gives rise to  another
{\em more abelian} algebra by making  some  structure constants vanish 
\cite{IWa,Segal,Saletan,MooPat,MonPat,Weimara,Gilmore},
while a {\em deformation} goes to another {\em less
abelian} algebra `producing' some non-zero  structure constants
\cite{Gilmore,gersten,Rosen,LevyNahas, nijen, BayenEtAl, WB,expansion}. The idea of contractions of Lie algebras and groups historically appeared  in relation with the non-relativistic limit which brings relativistic
mechanics (Poincar\'e group) to classical mechanics (Galilei group) \cite{IWa, Segal, LevyLeblond}.
In the framework of kinematical algebras the
scheme of contractions  is well known~\cite{BLL,Bacry}: starting from the
(anti)de Sitter algebra  a sequence of different contractions leads to
Poincar\'e, Newton--Hooke, Galilei, \dots, ending up at the last stage
in the so called Static algebra. From the viewpoint of the graded
contraction theory 
\cite{MooPat,MonPat}, the $(3+1)$D case has been 
studied in~\cite{Tolar}, and kinematical contractions in
arbitrary dimension have been obtained in~\cite{solgen}.
 
The theory of  deformations ~\cite{gersten, LevyNahas, BayenEtAl} has been mainly developed in connection with  cohomology techniques  which relate Lie algebra cohomology groups to Lie algebra deformations~\cite{nijen}. This procedure  
shows in a systematic way  which non-zero Lie brackets can be set different from
zero for a given Lie algebra leading to all its possible deformations. By
following this   approach   all the $(3+1)$D Galilean deformations were
obtained  in~\cite{figue}. Recently,  classical deformations (within this
type of procedures)  and contractions have been studied   altogether
in~\cite{montigny} (see also references therein).

In another approach to
deformations, one tries to obtain explicit realizations of a
deformed Lie algebra  in terms of the universal enveloping algebra of the initial
one~\cite{Gilmore,Rosen}, but as yet these   
have not  been fully systematized and   there is no
general constructive theory. The procedure given in~\cite{Gilmore,Rosen} provides
the deformations from the universal enveloping algebra of the inhomogeneous
pseudo-orthogonal algebra
$\mathfrak{iso}(p,q)$ to the semisimple ones $\mathfrak{so}(p+1,q)$ (with $p+q=N$); these contain  as
particular cases those deformations starting from the Euclidean  algebra and
leading to either the elliptic or hyperbolic ones as well as the
deformations from Poincar\'e to both de Sitter algebras.
A different method   \cite{WB}
enables one to perform  the  deformations $\mathfrak{t}_{qp}(\mathfrak{so}(p)\oplus
\mathfrak{so}(q))\to \mathfrak{so}(p,q)$. Similar deformations for unitary algebras can
be found in the above references.

  In~\cite{expansion}  we proposed another  
deformation method which was   formulated in
algebraic terms and based, like the above procedures, on the universal
enveloping algebra of initial Lie algebras. Nevertheless, the ideas  of~\cite{expansion} follow from geometrical considerations which require one to
 control  the deformation by a parameter which has a precise meaning: it is
the curvature of some (family of) homogeneous spaces  associated with the
initial and the deformed Lie algebras. The procedure rests on their
(quadratic) Casimir operators. This method  was applied to   the
deformations within the set of  $(2+1)$D kinematical
algebras~\cite{BLL} where the Lie algebras are contractions of
$\mathfrak{so}(4)$. In arbitrary dimension,  $\mathfrak{so}(N+1)$ and its kinematical contractions have as many Casimirs as
its rank~\cite{casimir}; only one is quadratic and the others are higher order polynomials in the generators. Then
it could happen that a method working for $\mathfrak{so}(4)$ (the only particular case where the  two Casimirs are quadratic) may not be directly
extendible to higher dimensions.

In this paper we   propose  a more general deformation
procedure, again based  on geometrical ideas (curvatures of
homogeneous spaces and Casimirs) which, in principle,  can be applied to any
Lie algebra in any dimension (not only a kinematical one).  Such a method,
presented in the next section, leads naturally to some operators fulfilling the deformed commutation rules as in~\cite{nijen,figue,montigny}, but it also provides
 explicit expressions of the deformed generators as elements of the
universal enveloping algebra of the initial Lie algebra. Thus the
resulting expressions may allow a further study of deformations for both operators and representations.

In particular, we apply this procedure to the
Galilei algebra in the realistic kinematical dimension, the $(3+1)$D
case. Hence the  basics of the Galilei algebra structure and its
main associated symmetric homogeneous spaces (spacetime and phase space) are
recalled in section 3. 
 This allows us to produce, in an explicit constructive way,
{\em two} physically relevant deformations. One reverses the
well known  non-relativistic limit  
by obtaining the   
Poincar\'e algebra expressed through the universal enveloping
Galilei one.  Furthermore  some Poincar\'e operators such as 
Pauli--Lubanski components,   angular momentum,  
position and velocity operators, etc.\  are
expressed in terms of Galilean ones. We remark that these deformed objects are not provided by cohomological procedures as in the
Galilean deformations studied in~\cite{figue}. The other deformation reverses the
ordinary zero-curvature limit in the non-relativistic Newton--Hooke 
spacetimes, whose generators are obtained within the
universal enveloping    
algebra of the centrally extended Galilei algebra.  Deformations of representations for both Newton--Hooke algebras
are also presented. These two deformations   are explicitly  developed in
sections 4 and  5, respectively.   

Our approach also cover some known cases;  for example, the deformations starting 
from the Poincar\'e algebra and leading to both (anti)de Sitter  algebras~\cite{Gilmore,Rosen}
and additionally yielding explicitly (anti)de Sitter representations from Poincar\'e
ones. This is described in section 6. Finally, some remarks and relationships with
quantum deformations are pointed out in section 7.


\sect{A   classical deformation method}

The Lie algebra deformation method proposed in
 \cite{expansion} for the  $(2+1)$D kinematical algebras~\cite{BLL,Bacry}
will be extended here in order to apply to any Lie algebra of  arbitrary dimension.

Let $\mathfrak{g}$ be a Lie algebra which is obtained through a contraction from
another Lie algebra $\mathfrak{g}'$, $\mathfrak{g}'\to \mathfrak{g}$, with Lie groups $G$ and $G'$. 
Suppose
that the contraction corresponds to the  vanishing of   the {\em
curvature} $\k$ of some non-flat symmetrical  homogeneous space $G'/H'$   and
thus leads to a {\em flat} space  $G/H$, where 
$H$ and $H'$ are the  Lie subgroups  of the (common)
subalgebras invariant  under the
contraction, and where both homogeneous spaces are of the same 
 {\em rank}  $\rangospace$ (defined by       the number of
independent invariants of a pair of points in the space under the action of
the group~\cite{Gilmore}; this definition holds for spaces
associated to both semisimple and non-semisimple Lie groups; see~\cite{Jordan} for the Euclidean space).

Next assume  that both  $\mathfrak{g}'$ and $\mathfrak{g}$ have
$\numcas$ functionally independent  Casimirs  ($\numcas\ge \rangospace$). We
shall denote by ${\cal C}_l$ $(l=1,\dots,\numcas)$ a linear basis of the
center of the universal enveloping algebra of the  contracted Lie algebra
$\mathfrak{g}$ (with $\k=0$) and by ${\cal C}'_l$ the corresponding one for the  deformed
Lie algebra $\mathfrak{g}'$ (with $\k\neq 0$); the numbering $(l=1,\dots,\numcas)$ in
the set of Casimirs corresponds to increasing order of the ${\cal C}_l$ as polynomials in the generators,  starting  from the quadratic one ${\cal C}_1$
(related to the Killling--Cartan form). Although the process involves the
choice of a basis, the end result will turn out to be independent of this
choice.   The deformation process starts  from the universal enveloping 
algebra $U(\mathfrak{g})$ of    $\mathfrak{g}$ and looks for a deformed algebra
$\mathfrak{g}'$ by the following    four steps:

\begin{itemize}

\item[(1)]
 Writing the first $\rangospace$ deformed Casimirs as  polynomials in
the  curvature $\k$ we aim to recover:
\be
{\cal C}'_s = {\cal C}_s +\k {\cal J}_s^{(1)} +
\k^2 {\cal J}_s^{(2)}+
\k^3 {\cal J}_s^{(3)} +\dots  \qquad s=1,\dots, \rangospace 
\label{bbba}
\ee
where ${\cal
C}_s$, ${\cal J}_s^{(k)}$   are {\em
independent} of $\k$; the ${\cal C}_s$ are just the contracted
Casimirs.

\item[(2)]
Working  in  $U(\mathfrak{g})$   within an irreducible
representation,    consider as  a `deformation seed' some linear
combination of the  $\rangospace$ terms {\em linear} in
the curvature $\k$:
\be
{\cal J}=\sum_{s=1}^{\rangospace} \alpha_s {\cal J}_s^{(1)}  
\label{bbbb}
\ee
where $\alpha_s$ are  constants to be  determined. Notice that any (linear,
compatible with the numbering criterion) change in the  initial choice of
the ${\cal C}'_s$ will lead to the same `seed' because each of the terms
linear in $\k$ in the new ${\cal C}'_s$ up to $\rangospace$-th  are
themselves a linear combination of the former; this means that the method is
intrinsic, even if it is formulated using a particular basis of the center
of the universal enveloping algebra.

\item[(3)]
For each initial generator $X_i$ of $\mathfrak{g}$, define  the deformed generators
$X'_i$ of $\mathfrak{g}'$  as  the elements in $U(\mathfrak{g})$ given by:
\be
X'_i:=\left\{\begin{array}{ll}
X_i &\  \mbox{if}\quad \comm{\cal J}{X_i} =0   \cr
\comm{\cal J}{X_i} &\ \mbox{if}\quad \comm{\cal J}{X_i}
\ne 0  
\end{array}
\right. . 
\label{bbbc}
\ee

\item[(4)]
Enforce on the  new generators $X'_i$ the condition that they span a Lie 
algebra isomorphic to $\mathfrak{g}'$. If this is possible  at all, this will determine
the constants $\alpha_s$.

\end{itemize}

Therefore this four-step procedure firstly lies in the element ${\cal J}$ (\ref{bbbb}) which is  formed by all the terms   `missing' in  $U(\mathfrak{g})$ (according to the rank and the particular homogeneous space chosen for each particular deformation)  in order to complete a base of  invariants for  $U(\mathfrak{g}')$. And secondly in the expression (\ref{bbbc}) which means that  the new (deformed) generators directly come  from such  a `missing' part in  $U(\mathfrak{g})$, whilst the unchanged ones are those commuting with ${\cal J}$  remaining in the same form as in $U(\mathfrak{g})$.
 
 We  stress that the resulting expressions  (\ref{bbbc}) may allow one to
obtain  operators and realizations of the deformed algebra $\mathfrak{g}'$  in terms of
those of  $U(\mathfrak{g})$  providing   deformations for Lie algebra representations of
the initial algebra.

It  is worth noticing a main difference of this approach with
other proposals~\cite{Gilmore,Rosen,WB}: here not only  the second-order 
${\cal C}_1$ but {\em all}  the Casimir invariants play a role in the
deformation process, and the end result is intrinsic and would not change
if a different choice for the Casimirs was used.  In particular, we shall
show that the fourth-order Pauli--Lubanski invariant   is {\em essential} in
order to recover the Poincar\'e algebra and some representations  from the Galilei case. The need of using the higher order
Casimirs might explain why the inverse process to  the In\"on\"u--Wigner
non-relativistic contraction has  remained, as a matter of fact,  as an
unsolved problem in the framework of operators/representations. In this
respect we recall that it was already clear in the 70's that higher-order
Casimirs (and not only the quadratic one) would be involved in higher-rank
deformations. This is explicitly stated by Gilmore~\cite{Gilmore}, but
apparently  this line was not pursued enough.


\sect{The Galilei  algebra and associated homogeneous spaces}

In order to apply the above method to the $(3+1)$D Galilei
algebra, $\gal\equiv \mathfrak{iiso}(3)$, we first recall some basics.  Let $H$,
$P_i$, $K_i$ and $J_i$  be the usual generators of Galilean time
translation, space translations, boosts and spatial rotations,
respectively, with   Lie brackets     given by
\be
\begin{array}{lll}
\comm{J_i}{J_j}=\varepsilon_{ijk}J_k & \qquad
\comm{J_i}{P_j}=\varepsilon_{ijk}P_k &
\qquad \comm{J_i}{K_j}=\varepsilon_{ijk}K_k \cr
\comm{P_i}{P_j}=0 &\qquad \comm{P_i}{K_j}=0 &
\qquad \comm{K_i}{K_j}=0 \cr
\comm{H}{P_i}=0 &\qquad \comm{H}{K_i}=-P_i &\qquad \comm{H}{J_i}=0 
\end{array}
\label{aa}
\ee
where $i,j,k=1,2,3$; sums over repeated indices will be generally  assumed from
now on. Hereafter we consider the $3+1$ notation and any 3-`vector'
  is
 denoted as  $\>X=(X_1,X_2,X_3)$; its  square and its scalar and cross
products   with other   vector, say
$\>Y=(Y_1,Y_2,Y_3)$, are
\be
\begin{array}{l}
\>X^2=X_1^2+X_2^2+X_3^2\qquad |\>X|=(X_1^2+X_2^2+X_3^2)^{1/2}\\[2pt]
\>X\producto\>Y=X_1Y_1+X_2Y_2+X_3Y_3\qquad
\>X\wedge \>Y=\>Z \quad \ Z_i= \varepsilon_{ijk}X_jY_k.
\end{array}
\label{aax}
\ee
Note that  $\>X$ and $\>Y$ are generically non-commutative objects so that
the ordering in these products must be preserved.

The Galilei algebra $\gal$ has two Casimir invariants (quadratic and quartic)
which read:
 \be 
 \ppa=\>P^2   \qquad
\ppb=\>W^2  
 \qquad \>W=\>K\wedge\>P   
\label{ab}
\ee
  corresponding  to the square of
the linear momentum $\>P$ (non-relativistic kinetic energy) and to  the
square of the angular momentum $\>W$, respectively.

Further computations within the
universal enveloping Galilei algebra  require use of   the  
identities:
\be
\>P\producto \>W  =0 \qquad
\>K\producto \>W  =0 
\label{apenda}
\ee
which imply that
\be
|\>P\wedge
\>W| =  |\>P||\>W|\qquad  |\>K\wedge
\>W| =  |\>K||\>W| .
\label{apendaa}
\ee
We shall also need the following   Lie brackets between  the Galilei
generators,  the angular momentum components $\ww{i}$ and the
products
$\>J\producto\>P$ and $\>J\producto\>W$:
\be
\begin{array}{l}
\comm{\ww{i}}{H}=\comm{\ww{i}}{P_j}= \comm{\ww{i}}{K_j}= 
\comm{\ww{i}}{\ww{j}}=0 \qquad  \comm{\ww{i}}{J_j}=\varepsilon_{ijk}\ww{k}
 \end{array}
\label{apendd}
\ee 
\be
\begin{array}{l}
\comm{\>J\producto\>P}{H}=  \comm{\>J\producto\>P}{P_i}= \comm{\>J\producto\>P}{J_i}=
0  \qquad  \comm{\>J\producto\>P}{K_i}=\ww{i}   \\[2pt]
\comm{\>J\producto\>P}{\ww{i}}=-\varepsilon_{ijk}  P_j\ww{k}\qquad  
\comm{\>J\producto\>W}{H}=  \comm{\>J\producto\>W}{J_i}=
\comm{\>J\producto\>W}{\ww{i}}=0   \\[2pt]
 \comm{\>J\producto\>W}{P_i}= \varepsilon_{ijk}   P_j\ww{k} \quad\   
\comm{\>J\producto\>W}{K_i}= \varepsilon_{ijk}  K_j\ww{k}  \quad\  
\comm{\>J\producto\>W}{\>J\producto\>P}=\varepsilon_{ijk}  J_i P_j\ww{k}  .
\end{array}
\label{apendf}
\ee 
Other   commutators which involve the Casimir invariants are given by:
\be
\begin{array}{ll}
\comm{\>J\producto\>P}{\varepsilon_{ijk}  P_j\ww{k}  }=\ppa\ww{i} &\qquad
\comm{\>J\producto\>P}{\varepsilon_{ijk}  K_j\ww{k} }=\>K\producto\>P\,
\ww{i}\\[2pt] \comm{\>J\producto\>W}{\varepsilon_{ijk}  P_j\ww{k}}=-\ppb
P_i &\qquad \comm{\>J\producto\>W}{\varepsilon_{ijk}  K_j\ww{k}}=-\ppb
K_i .
\end{array}
\label{apendh}
\ee

The   involutive automorphisms parity 
$\parity$ and the  product $\parity\timereversal$~\cite{BLL} (where $\timereversal$ is the
time-reversal)
\be
\begin{array}{ll}
\parity\timereversal: \quad &(H,\>P,\>K,\>J)\to 
(-H,-\>P,\>K,\>J) \cr
\parity: \quad &(H,\>P,\>K,\>J)\to  (H,-\>P,-\>K,\>J)
\end{array} 
\label{aae}
\ee
determine two Cartan decompositions of the Galilei Lie algebra $\gal$
\be
\begin{array}{llll}
\parity\timereversal: \quad &\gal=\mathfrak{p}^{(1)}\oplus \mathfrak{h}^{(1)} &\quad
\mathfrak{p}^{(1)}=\langle H,\>P\rangle &\quad
\mathfrak{h}^{(1)}=\langle  \>K,\>J\rangle \cr
\parity: \quad &\gal=\mathfrak{p}^{(2)}\oplus \mathfrak{h}^{(2)} &\quad
\mathfrak{p}^{(2)}=\langle \>P,\>K\rangle &\quad
\mathfrak{h}^{(2)}=\langle
H,\>J\rangle= \langle
H\rangle \oplus \langle \>J\rangle 
\end{array} 
\label{ae}
\ee
fulfilling
\be
\comm{\mathfrak{h}^{(l)}}{\mathfrak{h}^{(l)}}\subset \mathfrak{h}^{(l)} \qquad
\comm{\mathfrak{h}^{(l)}}{\mathfrak{p}^{(l)}}\subset \mathfrak{p}^{(l)} \qquad \comm{\mathfrak{p}^{(l)}}{\mathfrak{p}^{(l)}}=0 
\qquad l=1,2 .
\label{af}  
\ee 
 Consequently,  the Galilei group
$G\equiv IISO(3)$  is  the motion group of the two following 
symmetrical homogeneous spaces with dimension $D_l$, curvature $\k_l$
and rank $\rangospace_l$:
$$
\begin{array}{llll}
{\cal S}^{(1)}=G/H^{(1)}=IISO(3)/ISO(3) &\quad
D_1=4  &\quad
\k_1=0 &\quad  \rangospace_1=1 \\[2pt]
 {\cal S}^{(2)}=G/H^{(2)}=IISO(3)/(\R\otimes
SO(3)) &\quad
D_2=6 &\quad
\k_2=0 &\quad \rangospace_2=2.  
\end{array} 
$$

The Galilei subgroups   $H^{(1)}$, $H^{(2)}$  (whose Lie
algebras are $\mathfrak{h}^{(1)}$, $\mathfrak{h}^{(2)}$) are the stabilizer  subgroups
of an event  and of a time-like line, respectively. Therefore  
the  flat  rank-one space ${\cal S}^{(1)}$ is identified with the  
$(3+1)$D Galilean spacetime, while the flat rank-two space ${\cal
S}^{(2)}$ is the 6D space of time-like lines in the Galilean
spacetime  ${\cal S}^{(1)}$; we remark that ${\cal
S}^{(2)}$ itself   is the (classical) phase space of a
free Galilean particle. 
Hence this geometric approach makes two types of spaces to appear: spacetime, on one side, and phase spaces on the other. For a description of the structure of phase spaces behind the possible homogeneous spacetimes (Galilei, Newton--Hooke, Minkowski, (anti)de Sitter) as rank-two spaces, see~\cite{Jaca}. The phase spaces are the basic objects in the geometric quantization program, and through identification to orbits of the coadjoint action an interesting  link to the study and classification of representations,  the `orbit method', has been developed by Souriau, Kostant and Kirillov (for a recent overview of this method, see~\cite{KirillovLOT}). Our aim here is   more specific as we are not dealing with quantization at all.

Hence, as it is well known, two  possible
  deformations arise from Galilei in a natural way. 
First,  if
we consider   the Galilean
spacetime ${\cal S}^{(1)}$ we can
try  to reach a kinematics whose
associated spacetime has a {\em non-zero} curvature but keeps a
{\em flat} phase space   ${\cal S}^{(2)}$; that is,    we keep
non-relativistic (classical) mechanics.   The resulting symmetries span the two
Newton--Hooke (NH) algebras which are obtained through rank-one {\em
non-relativistic} deformations that  provide  {\em curved} spacetimes with
curvature  
$\k_1=\pm 1/\tau^2$ (where
$\tau$ is the universe `radius') but are still with {\em
absolute time}. The opposite process is the  spacetime
contraction  $\tau\to \infty$
   in the  NH algebras~\cite{BLL}.

 The second  possibility is to start with 
 the  rank-two space   ${\cal S}^{(2)}$ and introducing   a
(negative) curvature in 
${\cal S}^{(2)}$  as $\k_2=-1/c^2$, while keeping a flat spacetime ${\cal
S}^{(1)}$, and so reaching the Poincar\'e algebra. Hence  the usual (flat)
{\em phase space} supporting classical mechanics becomes the proper
`curved' phase space underlying relativistic mechanics~\cite{Jaca}.
Alternatively,  this process  can be seen as a
    relativistic deformation which  gives
rise to the Minkowskian spacetime, still a  {\em flat}
universe (spacetime) but now without {\em absolute time}. 
This is exactly the opposite
process to the well known non-relativistic limit or speed-space
contraction $c\to \infty$~\cite{IWa,Segal,Saletan,BLL}.

In the  next two sections we discuss in full explicit detail both types of
Lie algebra deformations together with a study of the corresponding
deformation of representations.


\sect{Poincar\'e from Galilei}

In this section we will see how the proposed deformation procedure leads to 
representations of the Poincar\'e algebra starting from those representations of the Galilei algebra with non-zero values for both Casimirs. These Galilei representations  will be termed here as `ordinary' just to emphasize that they are in some sense the `generic ones'. At the representation level the procedure requires to restrict oneself to those `ordinary' representations of the Galilei algebra to start with, and leads precisely to `non-ordinary' representations  of the Poincar\'e algebra, with both Casimirs vanishing.

The Lie brackets of the
$(3+1)$D Poincar\'e algebra $\poinc\equiv \mathfrak{iso}(3,1)$ read
\be
\begin{array}{lll}
\comm{J'_i}{J'_j}=\varepsilon_{ijk}J'_k & \qquad
\comm{J'_i}{P'_j}=\varepsilon_{ijk}P'_k  &
\qquad \comm{J'_i}{K'_j}=\varepsilon_{ijk}K'_k \\[2pt]
\comm{P'_i}{P'_j}=0 &\qquad \displaystyle{\comm{P'_i}{K'_j}=-\frac{1}{c^2}
\delta_{ij}H'}     &\qquad
\displaystyle{\comm{K'_i}{K'_j}=-\frac{1}{c^2}
\varepsilon_{ijk} J'_k} \\[4pt] 
\comm{H'}{P'_i}=0 &\qquad \comm{H'}{K'_i}=-P'_i  &\qquad \comm{H'}{J'_i}=0 . 
\end{array}
\label{ba}
\ee
 The two Poincar\'e Casimir invariants are given by
\be
\begin{array}{l}
 \displaystyle{{\cal C}'_1
=\>P'^2 -\frac 1{c^2} H'^2  \qquad
  {\cal C}'_2
=  \>{W'}^2-\frac 1{c^2} W'^2_0 } \\[2pt]
 \displaystyle{W'_0= \>J'\producto\>P' 
\qquad
 W'_i=-\frac 1{c^2} H' J'_i+\varepsilon_{ijk} K'_j P'_k .}
\end{array}
\label{bb}
\ee
For an elementary system, the Casimir ${\cal C}'_1$ is, up to a factor,  the square  of the energy-momentum 4-vector $(H',\>{P'})$; with the conventions here, this Casimir is equal to minus the system rest mass squared, while ${\cal C}'_2$ comes similarly from the square of the Pauli--Lubanski
4-vector $(W'_0,\>{W'})$, and is the square of the angular momentum in the rest frame; both quantities are the Poincar\'e invariants characterizing elementary systems. 

The Cartan decompositions  (\ref{ae}) also hold for the
Poincar\'e algebra, though in this case the vector subspaces
$\mathfrak{p}^{(l)}$  satisfy $\comm{\mathfrak{p}^{(1)}}{\mathfrak{p}^{(1)}}=0$ and
$\comm{\mathfrak{p}^{(2)}}{\mathfrak{p}^{(2)}}\subset \mathfrak{h}^{(2)}$ in the relations
(\ref{af}). Therefore   the Poincar\'e group  $P\equiv
ISO(3,1)$ is the motion group of the symmetric homogeneous
spaces defined by
$$
\begin{array}{llll}
 {\cal S}^{(1)}=P/H^{(1)}=ISO(3,1)/SO(3,1) &\ 
D_1=4  &\ 
\k_1=0 &\   \rangospace_1=1 \\[2pt]
{\cal S}^{(2)}=P/H^{(2)}=ISO(3,1)/(\R\otimes
SO(3)) &\ 
D_2=6 &\ 
\k_2=-1/c^2 &\  \rangospace_2=2 
\end{array}
\label{bbc}
$$
which are identified 
with the  $(3+1)$D Minkowskian spacetime   and the 6D (relativistic) phase
space   (or space of time-like lines), respectively.

The non-relativistic limit $c\to \infty$  leads to  the
contraction $\poinc\to \gal$ and makes equal to zero the curvature of
${\cal S}^{(2)}$, so that the commutators 
$\comm{P'_i}{K'_j}$, $\comm{K'_i}{K'_j}$ are equal to zero in $\gal$. Our aim
now is to reverse this process, that is, to try to recover $\poinc$  inside the
universal enveloping algebra of $\gal$,  $U(\gal)$, thus    introducing   the
relativistic constant $c$.


\subsect{The relativistic Galilei deformation}

As the intended deformation $U(\gal)\to \poinc$, which would produce the Poincar\'e algebra from the Galilei one, is a {\em rank-two deformation}, the first step of
the deformation method (\ref{bbba}) requires us to   rewrite {\em both} 
Poincar\'e Casimirs so as to explicitly display the powers of the
curvature
$-1/c^2$ of the phase space  ${\cal S}^{(2)}$; this gives the ${\cal J}_i^{(l)}$: 
\bea
&&\!\!\!\!\!\!\!\!\!\!\!\!\!\!\!\!\!
\displaystyle{{\cal C}'_1={\cal C}_1  -\frac 1{c^2}  {\cal
J}_1^{(1)}
\qquad {\cal J}_1^{(1)}=H^2  }\label{cf} 
\\
&&\!\!\!\!\!\!\!\!\!\!\!\!\!\!\!\!\!
\displaystyle{{\cal C}'_2={\cal C}_2 -\frac 1{c^2} {\cal
J}_2^{(1)} +
\left(-\frac 1{c^2}\right)^2\!\! {\cal J}_2^{(2)}  \quad\ 
{\cal J}_2^{(1)}=2 H\,\>J\producto\>W + (\>J\producto\>P)^2 \quad\ 
{\cal J}_2^{(2)}=H^2\,\>J^2  }
\nonumber
\eea
where ${\cal C}_1$, ${\cal C}_2$ are  the {\em Galilei} Casimirs
(\ref{ab}). We now consider 
 the linear combination (\ref{bbbb})
\be
{\cal J}=\alpha_1 {\cal J}_1^{(1)} + \alpha_2{\cal J}_2^{(1)}=
\alpha_1 H^2 + \alpha_2\left( 2   H\,\>J\producto\>W + 
(\>J\producto\>P)^2 \right)
\label{ch}
\ee
to be used as a seed to provide generators  $X'_k$  intended to close a Poincar\'e algebra.
These are the elements of $U(\gal)$ obtained by following  the third step (\ref{bbbc}),
which   requires commuting ${\cal J}$ with the Galilei generators $X_k$: 
\be
\begin{array}{l}
H'=H \qquad J'_i=J_i \qquad
P'_i=2\AA_2 H  \varepsilon_{ijk}  P_j\ww{k}   \\[2pt]
 K'_i=-2\AA_1 H P_i+2\AA_2 \left(
\>J\producto\>P\,\ww{i} - \>J\producto\>W\, P_i  
+   H \varepsilon_{ijk} K_j \ww{k}   + \frac 32
\varepsilon_{ijk} P_j \ww{k} \right). 
\end{array} 
\label{bg} 
\ee
  The last step is to enforce the deformed generators (\ref{bg}) to
fulfil the Lie brackets of the Poincar\'e algebra  (\ref{ba});
the final result is established by:
\medskip

\noindent
{\bf Proposition 1.} {\em For a given irreducible representation of the Galilei algebra $\gal$ with non-vanishing Casimirs ${\cal C}_1\neq0, {\cal C}_2\neq0$, the gene\-rators defined by   (\ref{bg})
close the $(3+1)$D  Poincar\'e  algebra $\poinc$ whenever  the constants
$\AA_1$, $\AA_2$ are related to the Casimirs  values by 
\be
\AA_1 \ppa + \AA_2\ppb=0 \qquad
\AA_2^2 = \frac 1{4 c^2 \ppa\ppb}. 
\label{bh}
\ee
}

\medskip
Notice that this process will only work for those irreducible representations of the starting Galilei algebra with both Casimirs {\it different from zero}. There is no a standard naming for these representations, but as this condition is essential in our approach, we will refer here to these as `ordinary' Galilei representations. 

The proof of this result is a direct computation. Some generators stay unchanged in this deformation and span the   subalgebra   
$\mathfrak{h}^{(2)}=\langle H,\>J\rangle$.  By taking into account
the second  Cartan decomposition (\ref{ae}) it can be checked
that the assumptions of the  proposition 1 of   \cite{expansion}
are automatically fulfilled,  which   implies that the 
Lie brackets $\comm{\mathfrak{h}^{(2)}}{\mathfrak{h}^{(2)}}$ and $\comm{\mathfrak{h}^{(2)}}{\mathfrak{p}^{(2)}}$  remain in
the same form as in the Galilei algebra, as it should be for 
Poincar\'e. Therefore,  we have only to compute the   commutators
involving  generators in $\mathfrak{p}^{(2)}$:  $\comm{P'_i}{P'_j}$,   
$\comm{P'_i}{K'_j}$ and
$\comm{K'_i}{K'_j}$.

As it is shown  in (\ref{apendd}), $H$, $P_i$ and $W_j$
commute amongst themselves, so that  
$\comm{P'_i}{P'_j}=0$. Let us compute  
$\comm{P'_i}{K'_l}$; by using (\ref{apendd}) and (\ref{apendf})  we obtain
that
\bea
 \comm{P'_i}{K'_l}&\!\!\!=\!\!\!& -4 \AA_2^2\, \varepsilon_{ijk}
H\comm{P_j}{\>J\producto\>W} P_l\ww{k}+4 \AA_2^2 \,\varepsilon_{ijk} H P_j\comm{\ww{k}}{\>J\producto\>P}\ww{l}
\nonumber\\[2pt]
&& \qquad +4 \AA_2^2\, \varepsilon_{ijk}\varepsilon_{lmn} 
H \comm{H}{K_m}  P_j \ww{k}\ww{n} \nonumber\\[2pt]
&\!\!\!=\!\!\!& 4\AA_2^2\,H \varepsilon_{ijk}
\left(\varepsilon_{jmn}  P_l\ww{k}
+\varepsilon_{kmn}   P_j\ww{l}
-\varepsilon_{lmn}   P_j\ww{k} \right)P_m\ww{n} .
\nonumber
\eea
Thus, if $i\ne l$,   $\comm{P'_i}{K'_l}=0$, but whenever $i=l$ the
Galilei Casimirs (\ref{ab}) can be recovered in the commutator as
$$
\comm{P'_i}{K'_i}=-4 \AA_2^2 H \>P^2 \>W^2= -4 \AA_2^2 H  \ppa\ppb =
-4 \AA_2^2 \ppa\ppb H' 
$$
and this will coincide with the expected Poincar\'e commutation relation whenever
$$
-4 \AA_2^2 \ppa\ppb = -\frac{1}{c^2}
$$
which leads to  the second relation of (\ref{bh}).  
Next,  let us compute
\bea
\!\!\!\!\!\!\!\!\!
 \comm{K'_i}{K'_l}&\!\!\!=\!\!\!&
8\AA_1\AA_2 H (\varepsilon_{ljk}P_i-\varepsilon_{ijk}P_l)P_j\ww{k}
  -4 \AA_2^2 H \ppb\,\varepsilon_{ilk}\ww{k} \nonumber\\[2pt]
&&- 4 \AA_2^2 H\varepsilon_{ijk}\varepsilon_{lmn}\varepsilon_{sjm}
\ww{n}\ww{s}\ww{k} -4 \AA_2^2 \comm{\>J\producto\>W}{\>J\producto\>P} 
(P_i\ww{l} -P_{l}\ww{i})\nonumber\\[2pt] && 
\displaystyle{+ 4 \AA_2^2 \left\{\>J\producto\>W 
(\varepsilon_{ljk}P_i-\varepsilon_{ijk}P_l)P_j\ww{k}
+ \>J\producto\>P  
(\varepsilon_{ljk}\ww{i}-\varepsilon_{ijk}\ww{l})\ww{j}P_k\right\} }.
\nonumber
\eea
By using identities  (\ref{apenda}), simplifying, grouping
terms   in order to  construct the Galilei Casimirs and using the recently obtained relation  we find that
\bea
 \comm{K'_i}{K'_l}&\!\!\!=\!\!\!&
-8 \AA_2 H \varepsilon_{ilk} (\AA_1\ppa+ \AA_2\ppb )\ww{k}
- 4 \AA_2^2 \varepsilon_{ilk} (\>J\producto\>W \,\ppa\ww{k}
+\>J\producto\>P \,\ppb P_k)
\nonumber\\[2pt]
&&  
-4 \AA_2^2\, \varepsilon_{mns}J_mP_n\ww{s}  
(P_i\ww{l} -P_{l}\ww{i})\nonumber\\[2pt]
 &\!\!\!=\!\!\!&-8 \AA_2 H  \varepsilon_{ilk}\left\{\AA_1\ppa
+ \AA_2\ppb \right\}\ww{k}
-4 \AA_2^2  \varepsilon_{ilk} J_k \ppa\ppb \nonumber\\
 &\!\!\!=\!\!\!&-8 \AA_2 H  \varepsilon_{ilk}\left\{\AA_1\ppa
+ \AA_2\ppb \right\}\ww{k}
-\frac 1{c^2}
\varepsilon_{ilk} J'_k
\nonumber
\eea
so this coincides with the Poincar\'e commutator whenever
$$
 \AA_1\ppa + \AA_2\ppb   = 0.
$$
Hence proposition 1 is proven.

Consequently, this Galilean
deformation  reverses the non-relativistic contraction of the
Poincar\'e algebra   by  {\em introducing} a 
negative curvature $\k_2=-1/c^2$ in the 6D Galilean phase space  
${\cal S}^{(2)}$.   Recall   that the space ${\cal
S}^{(2)}$ is rank-two, so its geometry is {\em very} different from a
rank-one space, either   curved or  
flat~\cite{Jaca}.    At the level of the Cartan
decompositions the deformation gives
$\comm{\mathfrak{p}^{(2)}}{\mathfrak{p}^{(2)}}=0\to \comm{\mathfrak{p}^{(2)}}{\mathfrak{p}^{(2)}}\subset 
\mathfrak{h}^{(2)}$.


\subsect{Relativistic Poincar\'e operators through non-relativistic Galilei
ones}

Once the constants $\alpha_1$ and $\alpha_2$ have been determined, the 
expressions of the Poincar\'e generators in terms  of those of the universal envolvent algebra $U(\gal)$ (\ref{bg}) of the Galilei algebra can be recast very compactly in  vector form. We restrict from the beginning to `ordinary' representations of the Galilei algebra, with ${\cal C}_1\neq0$ and ${\cal C}_2\neq0$. Choose the positive root determination for $\alpha_2$, and use the notation (\ref{aax}), writing $\ppa^{1/2}=|\>P|$ and $\ppb^{1/2}=|\>W|$. For the deformed generators we obtain:
\be
\begin{array}{l}
\displaystyle{c\>{P'}= H\,\frac{\>P\wedge \>W}{|\>P||\>W|
}    \qquad H' = H\qquad   \>J' = \>J} \\[12pt]
\displaystyle{c\>{K'}= H\left(
\frac{|\>W|}{\>P^2}\,\frac{\>P}{|\>P|}+ 
\frac{\>K\wedge \>W}{|\>P||\>W| }  \right)+
\left( \frac{\>J\producto\>P}{|\>P|}\,\frac{\>W}{|\>W|}-
\frac{\>J\producto\>W}{|\>W|}\,\frac{\>P}{|\>P|}    
\right)+
\frac 32\,\frac{\>P\wedge \>W}{|\>P||\>W|}  } \\[12pt]
\displaystyle{\qquad = \frac 1{|\>P||\>W| }  \left\{
H\left(
\frac{\>W^2}{\>P^2}\, {\>P} + 
 {\>K\wedge \>W} 
\right)- {\>J\wedge(\>P\wedge \>W)}  +
\frac 32\, {\>P\wedge \>W}  \right\} }.
 \end{array} 
\label{bi} 
\ee

Next we can deduce  Poincar\'e operators as elements in $U(\gal)$.
By using the relations (\ref{apenda})--(\ref{apendh}) we find that
\be
c^2 \>P'^2=H'^2\qquad
c^2 \>{K'}\wedge \>{P'}= H\>J-H\left\{\frac{\>J\producto(\>P\wedge
\>W)}{\>P^2\>W^2} \right\}{\>P\wedge \>W}  .
\label{bbii}
\ee
Thus 
the Pauli--Lubanski components turn out  to be
\be
W'_0=\frac 1 c \, H\>J\producto\frac{(\>P\wedge \>W)}{|\>P||\>W|}
\qquad
\>W'=-\frac 1{c^2}\, H \left\{\frac{\>J\producto(\>P\wedge
\>W)}{|\>P||\>W|}\right\}\frac{\>P\wedge \>W}{|\>P||\>W|}  .
\label{bj}
\ee
These expressions explicitly show how the   component 
$W'_0$, not appearing in the Galilei case (\ref{ab}), can be constructed
and, moreover, they lead to
\be
\>W'= -\frac{1}{c}\,W'_0\,\frac{\>P\wedge \>W}{|\>P||\>W|} \qquad c^2
\>W'^2=W_0'^2.
\ee
If  we now substitute these expresions into the Poincar\'e Casimirs
${\cal C}'_1, {\cal C}'_2$ given in (\ref{bb}) it can be checked that both of them vanish. This, in turn, means that  under this deformation {\em any irreducible Galilean representation with both  ${\cal C}_1\ne 0,\  {\cal
C}_2\ne 0$} gives rise to a {\em massless} and {\em spinless}  Poincar\'e
representation with both ${\cal C}'_1=  {\cal C}'_2=0$. Hence, the expressions  (\ref{bi}) can be interpreted as a correspondence between non-relativistic `ordinary' systems  and  relativistic zero-mass systems. This suggests some kind of relationship between   these operators   and  the null-plane (or light-cone) Poincar\'e framework~\cite{leutwyler}.

Furthermore, a remarkable fact is the   appearance in (\ref{bi}) of
a Galilean angular momentum~\cite{Bacrybis} 
\be
\lambda_p=  \frac{\>J\producto\>P}{|\>P|}=\>J\producto{\>u}_p  
\label{bk}
\ee
which generates an $\mathfrak{so}(2)$ Lie algebra and  
corresponds to the projection of $\>J$ along the
direction ${\>u}_p$ of the motion $\>P$; recall that we have required $\ppa^{1/2}=|\>P|\ne 0$. Similarly we can consider two other helicity-type
operators   defined along two orthogonal directions to $\>P$ determined by
$\>W=\>K\wedge\>P  $ and
$\>P\wedge\>W$ (note also that accordingly $\ppb^{1/2}=|\>W|$ has been assumed to be different from zero):
\be
\lambda_w :=  \frac{\>J\producto\>W}{|\>W|}=\>J\producto{\>u}_w 
\qquad
\lambda_{pw} := \frac{\>J\producto(\>P\wedge
\>W)}{|\>P\wedge\>W|}=\>J\producto{\>u}_{pw}  .
\label{bl}
\ee
By taking into account (\ref{apendf}) and (\ref{apendh}), we find
that these three Galilean `angular momenta' close an $\mathfrak{so}(3)$ Lie algebra:
\be
\comm{\lambda_w}{\lambda_p}=\lambda_{pw}\qquad
\comm{\lambda_{pw}}{\lambda_w}=\lambda_{p}\qquad
\comm{\lambda_p}{\lambda_{pw}}=\lambda_{w} .
\label{bm}
\ee
In this way, we can  write
the  Poincar\'e angular momentum $\lambda'$  and the Pauli--Lubanski
components   in terms of the Galilean `helicity'  $\lambda_{pw}$ in the form
\be
\lambda'=c\,\frac{W'_0}{H'}= \frac{\>J'\producto\>P'}{|\>P'|}\equiv
\lambda_{pw}
\qquad  W'_0=\frac 1 c \, H \,\lambda_{pw} 
\qquad
\>W'=-\frac 1{c^2}\, H \, \lambda_{pw}\,{\>u}_{pw}  .
\label{bn}
\ee

Now we consider the  Poincar\'e
operators
$Q'_i$ defined by Bacry~\cite{Bacryposa} (of course, for $H'\neq0$):
\be
Q'_i=\frac{c^2}{2}\left( \frac 1 {H'} \, K'_i + K'_i \, \frac 1 {H'}\right).
\label{bo}
\ee
By introducing (\ref{bi}) and applying the relations
(\ref{apenda})--(\ref{apendh}) we find that these operators are      
expressed through elements of
$U(\gal)$ as
\bea
&&\!\!\!\!\!\!\!\!
 \frac 1 c\,\>Q'= 
\frac{|\>W|}{\>P^2}\,\frac{\>P}{|\>P|}+ 
\frac{\>K\wedge \>W}{|\>P||\>W| }  +\frac 1 H
\left( \frac{\>J\producto\>P}{|\>P|}\,\frac{\>W}{|\>W|}-
\frac{\>J\producto\>W}{|\>W|}\,\frac{\>P}{|\>P|} +
 \frac{\>P\wedge \>W}{|\>P||\>W|}     \right) \nonumber\\
&&\quad   =
\frac{|\>W|}{\>P^2}\,{\>u}_p+ 
\frac{\>K\wedge \>W}{|\>P||\>W| }  +\frac 1 H
\left(\lambda_p {\>u}_w-
\lambda_w {\>u}_p +{\>u}_{pw}  \right) .
\label{bp}
\eea
Some expected properties fulfilled by $\>Q'$ are consistently recovered
under deformation. Working within $U(\gal)$ we obtain  
\be
\comm{J'_i}{Q'_j}=\varepsilon_{ijk} Q'_k \qquad \comm{Q'_i}{P'_j}=\delta_{ij} .
\ee
The former Lie bracket shows   that $\>Q'$ behaves as a 3-vector under
rotations $\>J'=\>J$, while the latter provides   the natural canonical  commutation relations between Galilean position and momenta operators; had we kept a definition for Galilean
position operators $Q_i$ analogous to (\ref{bo}) then we would had obtained the unwanted result  
$\comm{Q_i}{P_j}=0$. Secondly, cumbersome computations lead to
\bea
&&\comm{Q'_i}{Q'_j}=\frac{c^2}{H^2}\,\varepsilon_{ijk}\left(  
\frac{\>J\producto\>P}{|\>P|}\,\frac{P_k}{|\>P|}+
\frac{\>J\producto\>W}{|\>W|}\,\frac{W_k}{|\>W|}-J_k   \right)\nonumber\\
&&\qquad\qquad  =\frac{c^2}{H'^2}\left( Q'_i P'_j- Q'_j P'_i -
\varepsilon_{ijk}
 J'_k\right) .
\eea
As a byproduct, the Poincar\'e kinematical
observables  $\Sigma'_k=J'_k- 
\varepsilon_{ijk}Q'_i P'_j$ can be written in terms of Galilean operators as
\be
{\bf \Sigma'} =\>J   -
\frac{\>J\producto\>P}{|\>P|}\,\frac{\>P}{|\>P|}-
\frac{\>J\producto\>W}{|\>W|}\,\frac{\>W}{|\>W|}
=\>J   -\lambda_p {\>u}_{p} -\lambda_w {\>u}_{w} .
\ee
Finally, the Poincar\'e velocity operator $\>V'$ turns out to be
\be
\>V'=\comm{\>Q'}{H'}= c\, \frac{\>P\wedge
\>W}{|\>P||\>W|}= c^2
\, \frac{\>P'}{H'}=c \,    \frac{\>P'}{|\>P'|} 
\label{bq}
\ee
provided that $H'=c|\>P'|$ since this deformation gives rise to Poincar\'e
massless representations. The result (\ref{bq}) can be
interpreted as a relationship between  the motion of a 
non-relativistic particle  along the direction
${\>u}_{pw}$ with speed $c$  and   the motion
of a free massless relativistic particle  along the direction ${\>u}'_{p'}$.


\sect{Newton--Hooke from centrally extended 
Galilei}

Besides   Poincar\'e, the
Galilei algebra has two other kinematical neighbours:
the oscillating $\newhooke_+$ and the expanding  $\newhooke_-$ NH
algebras \cite{BLL}.  These are the Lie algebras of  the
motion groups of {\em absolute time} universes   with {\em non-zero
curvature} $\k_1=\pm 1/\tau^2$ where $\tau$ is the universe time
radius.    The commutation rules of  $\newhooke_\pm$
are given  by 
\be
\begin{array}{lll}
\comm{J'_i}{J'_j}=\varepsilon_{ijk}J'_k & \qquad
\comm{J'_i}{P'_j}=\varepsilon_{ijk}P'_k  &
\qquad \comm{J'_i}{K'_j}=\varepsilon_{ijk}K'_k \cr
\comm{P'_i}{P'_j}=0 &\qquad  \comm{P'_i}{K'_j}=0    &\qquad  \comm{K'_i}{K'_j}= 0
\cr  \comm{H'}{P'_i}=\k_1 K'_i &\qquad \comm{H'}{K'_i}=-P'_i  &\qquad
\comm{H'}{J'_i}=0   .
\end{array}
\label{ca}
\ee
The Casimirs of $\newhooke_\pm$ turn out to be:
\be
\begin{array}{l}
{\cal C}'_1
=\>P'^2 +\k_1 \>K'^2   \qquad 
{\cal C}'_2
=  \>{W'}^2  
\end{array}
\label{cb}
\ee
where the components of $\>{W'}$ are formally identical to the
Galilean ones (\ref{ab}).

Due to the non-zero Lie brackets $\comm{H'}{P'_i}=\k_1 K'_i$,   the
vector subspaces $\mathfrak{p}^{(l)}$ of the Cartan decompositions  
(\ref{ae}) now verify  $\comm{\mathfrak{p}^{(1)}}{\mathfrak{p}^{(1)}}\subset
\mathfrak{h}^{(1)}$ and $\comm{\mathfrak{p}^{(2)}}{\mathfrak{p}^{(2)}}=0$. The NH groups 
${N}_\pm$ are the motion groups  of the following  symmetric
homogeneous spaces 
$$
\begin{array}{ll}
 {\cal S}^{(1)}=N_+/H^{(1)}=T_6(SO(2)\otimes SO(3))/ISO(3) 
&\  \k_1=1/\tau^2 \cr
{\cal S}^{(2)}=N_+/H^{(2)}=T_6(SO(2)\otimes
SO(3))/(SO(2)\otimes SO(3)) & \ 
\k_2=0   \\[6pt]
 {\cal S}^{(1)}
=N_-/H^{(1)}=T_6(SO(1,1)\otimes SO(3))/ISO(3) &\ 
\k_1=-1/\tau^2 \cr
{\cal S}^{(2)}=N_-/H^{(2)}
=T_6(SO(1,1)\otimes SO(3))/(SO(1,1)\otimes
SO(3)) &\ 
\k_2=0   .
\end{array} 
\label{cd}
$$
The spaces  ${\cal S}^{(1)}$ and  ${\cal S}^{(2)}$ 
correspond, in this order, to the   $(3+1)$D non-relativistic
curved spacetime~\cite{Jaca,Aldrovandi}   and the 6D flat  phase space (or
space of time-like worldlines in the spacetime).

The limit $\k_1\to
0$ ($\tau\to \infty$)  produces the spacetime
contraction $\newhooke_\pm\to \gal$. We consider the
opposite process which would introduce a universe
time radius $\tau$  in  $U(\gal)$.  As we are
dealing with rank-one deformations we only consider the first Casimir 
(\ref{cb}), so that   the  `seed' element   (\ref{bbbb}) is simply $\jjj=
\AA_1 {\cal J}_1^{(1)} =  \AA_1 \>K^2$. The   
deformed generators (\ref{bbbc})  are $H'=2\AA_1 \>K\producto\>P$ and all the
remaining ones  unchanged. However
these new generators {\em do not span} $\newhooke_\pm$.
Hence it is necessary to take an initial Lie algebra  {\em less
abelian} in order to be able to perform the deformation. Thus  we start from
a central extension of $\gal$ (as in lower dimensions~\cite{expansion,nieto}), with central generator $\Xi$ and central charge $\ext$ (the mass of the
Galilean particle as a Galilean elementary system \cite{LevyLeblond}). The commutation rules of the centrally extended Galilei algebra
$\overline\gal$ are given by (\ref{aa}) once  the vanishing brackets
$\comm{K_i}{P_j}$ are replaced by  
\be
\comm{K_i}{P_j}=\delta_{ij}\ext\,\Xi \qquad
\comm{\Xi}{\cdot\,}=0 .
\ee
We now apply (\ref{bbbc}) with  $ \jjj=  \AA_1 \>K^2
$ for the Lie
brackets of $\overline\gal$,  finding     the  deformed
generators, which live on the universal enveloping algebra 
$U(\overline\gal)$:
\be 
  J'_i=J_i \qquad K'_i=K_i \qquad   
 H'=2\AA_1 \>K\producto\>P - 3 \AA_1  \ext \,\Xi \qquad
P'_i=2\AA_1  \ext\, \Xi\, K_i .
\label{cj}
\ee 
Then this deformation is characterized by:

\noindent
{\bf Proposition 2.} {\em The generators defined in terms of the (extended, with
$m\neq0$) Galilean ones by (\ref{cj})   span the $(3+1)$D  NH algebras   provided
that    the constant $\AA_1$ fulfils
\be
\AA_1^2 = -\frac {\k_1}{4 \ext^2 \Xi^2}
= \mp \frac {1}{4 \tau^2 \ext^2 \Xi^2} 
\qquad \mbox{for}\quad \newhooke_\pm .
\label{ck}
\ee}

\noindent
{\em Proof:}  The generators which are unchanged  in the
deformation close the isotopy subalgebra of an event $\mathfrak{h}^{(1)}= \langle
\>K,\>J\rangle$. Thus the   proposition 1 of   \cite{expansion} can be
applied and we only need to compute the   Lie brackets involving   generators
of  $\mathfrak{p}^{(1)}$. By direct computations we obtain the relation
(\ref{ck}):
$$
\comm{P'_i}{P'_j}=0 \qquad
\comm{H'}{P'_i}=-4\AA_1^2\ext^2\Xi^2 K_i\equiv \k_1 K'_i .
$$

Therefore  this deformation introduces a constant
 curvature $\k_1$ in the flat  spacetime
$G/{H}^{(1)}$, leading to  
curved spacetimes $N_\pm/{H}^{(1)}$ with the transition
$\comm{\mathfrak{p}^{(1)}}{\mathfrak{p}^{(1)}}=0\to \comm{\mathfrak{p}^{(1)}}{\mathfrak{p}^{(1)}}\subset \mathfrak{h}^{(1)}$.  All the phase spaces ${\cal S}^{(2)}$ remain flat.


\subsect{Deformation of Galilei representations}

If we set $\alpha_1=\sqrt{- \k_1}/({2\ext \Xi})$ and
substitute  in   the deformed generators (\ref{cj}), we find that these read
\be
 H'=\sqrt{- \k_1}\left( \frac{\>K\cdot\>P}{\ext\,\Xi}-\frac 32\right) 
\qquad
\>P'=\sqrt{- \k_1}\ \>K .
\label{reprenewtona}
\ee
Hence  ${\cal C}'_1=0$ and $\>{W'}=\>K'\wedge \>P'=0$. Consequently, 
any irreducible representation of $\overline\gal$ with $m\ne0$ gives
rise to a representation of $\newhooke_\pm$ with zero Casimirs eigenvalues  as both deformed ${\cal C}'_l$ (\ref{cb})  vanish.
 In particular, let us consider the  
representation of $\overline\gal$ in terms of the  momenta
$\>p=(p_1,p_2,p_3)$ given by Bacry~\cite{Bacrybis}:
\be
J_i=\varepsilon_{ijk}\,p_k\,\frac{\partial}{\partial p_j}+S_i 
\qquad K_i=m\,\frac{\partial}{\partial p_i} \qquad 
H=\frac{\>p^2}{2m}+a ,\qquad P_i=p_i \qquad \Xi=1
\ee
where $a$ is a constant  and
$S_i$ are spin operators fulfilling
$\comm{S_i}{S_j}=\varepsilon_{ijk}\,S_k$. Then the relations (\ref{reprenewtona})
lead  to the following representation for  the transformed
generators of 
$\newhooke_\pm$:
\be
  H'=\sqrt{- \k_1}
\left(\>p \cdot \frac{\partial}{\partial
\>p}+ \frac{3}{2} \right) \qquad
\>P'=\sqrt{- \k_1}\,  m \,\frac{\partial}{\partial \>p} 
 \label{reprenewtonb}
\ee
keeping $\>J'=\>J$ and $\>K'=\>K$.  


\sect{(Anti)de Sitter from Poincar\'e }

Known Lie algebra deformations   $\mathfrak{iso}(p,q)\to
\mathfrak{so}(p+1,q)$~\cite{Gilmore,Rosen} can   be recovered from our approach, which
also provides the explicit form of the operators representing the deformed generators
inside the initial universal enveloping algebra. As an example, in this section we give
either anti-de Sitter
$\desitter_+=\mathfrak{so}(3,2)$ or    de Sitter $\desitter_-=\mathfrak{so}(4,1)$
algebras and representations  as  deformations of Poincar\'e ones. To maintain
consistency with the notation in the previous part of the paper, we will denote here 
$X'$ any element of $U(\poinc)$ (as (\ref{ba}) and (\ref{bb})) and $X''$
any (anti)de Sitter object. The (anti)de Sitter commutation
relations   read
\be
\begin{array}{lll}
\comm{J''_i}{J''_j}=\varepsilon_{ijk}J''_k & \quad
\comm{J''_i}{P''_j}=\varepsilon_{ijk}P''_k  &
\quad \comm{J''_i}{K''_j}=\varepsilon_{ijk}K''_k \\[2pt]
\displaystyle{\comm{P''_i}{P''_j}=-\frac{\k_1}{c^2}\varepsilon_{ijk}J''_k}
&\quad
\displaystyle{\comm{P''_i}{K''_j}=-\frac{1}{c^2}
\delta_{ij}H''}     &\quad
\displaystyle{\comm{K''_i}{K''_j}=-\frac{1}{c^2}
\varepsilon_{ijk} J''_k} \\[4pt]
\comm{H''}{P''_i}=\k_1 K''_i &\quad \comm{H''}{K''_i}=-P''_i  &\quad
\comm{H''}{J''_i}=0  
\end{array}
\label{fa}
\ee
where $\k_1$ is the curvature of the spacetime: $\k_1=1/\tau^2$ for
$\desitter_+$ and  $\k_1=-1/\tau^2$ for $\desitter_-$. In both cases the 
constant $\tau$ plays the role of the curvature radius of the (anti)de Sitter universe;
we recall that in our conventions $\tau$ is dimensionally a time (so the `length
radius' would be $c\tau$) and the sign of spacetime curvature has been chosen so that
$\k_1$ is related to the acceleration of separation of {\it time-like} geodesics with
the same sign as in the Riemannian case (the sign is opposite for space-like
geodesics).  The  Casimir invariants are~\cite{casimir}:
\be
\begin{array}{l}
\displaystyle{{\cal C}''_1
=\>P''^2 -\frac 1{c^2} H''^2 +\k_1 \left( \>K''^2- \frac
1{c^2}\, \>J''^2\right) }\\[2pt] 
\displaystyle{{\cal C}''_2
=\>{W''}^2-\frac 1{c^2} W''^2_0
-\frac{\k_1}{c^2}\left(\>J''\producto\>K'' \right)^2}
\end{array}
\label{fb}
\ee
where $\>{W''}$,  $W''_0$ are formally identical to the
Poincar\'e Pauli--Lubanski components (\ref{bb}).
Now, both vector subspaces $\mathfrak{p}^{(l)}$ 
(\ref{ae})  fulfil $\comm{\mathfrak{p}^{(l)}}{\mathfrak{p}^{(l)}}\subset
\mathfrak{h}^{(l)}$, and  the  symmetric homogeneous (anti)de Sitter  spaces are
defined by
$$
\begin{array}{ll}
 {\cal S}^{(1)}=AdS/H^{(1)}= SO(3,2)/SO(3,1) 
&\quad  \k_1=1/\tau^2 \cr
{\cal S}^{(2)}=AdS/H^{(2)}=SO(3,2)/(SO(2)\otimes SO(3)) & \quad 
\k_2=-1/c^2   \\[6pt]
 {\cal S}^{(1)}
=dS/H^{(1)} =SO(4,1)/SO(3,1) 
&\quad  \k_1=-1/\tau^2 \cr
{\cal S}^{(2)}=dS/H^{(2)}=SO(4,1)/(SO(1,1)\otimes SO(3)) & \quad 
\k_2=-1/c^2    .
\end{array}
$$
Hence the spaces ${\cal S}^{(1)}$   are the curved relativistic spacetimes (with the two possible signs of the curvature), and  
the two `phase spaces' ${\cal S}^{(2)}$ are the corresponding spaces of
time-like lines; both are (negatively) curved phase spaces.

The deformation
$U(\poinc)\to \{\desitter_\pm\}$ is rank-one with underlying space ${\cal
S}^{(1)}$ and curvature $\k_1$. Then only the first Casimir ${\cal
C}''_1$ (\ref{fb}) determines the `seed' element (\ref{bbbb}):
\be
{\cal J}=\AA_1 \left( \>K''^2- \frac
1{c^2}\, \>J''^2\right)
\ee
so that the deformed generators, defined on $U(\poinc)$,   turn out to be
\bea
&&  J''_i=J'_i \qquad K''_i=K'_i \qquad   
 H''=\frac{2\AA_1}{c^2}\left( c^2\>K'\producto\>P' - \frac 3 2 H'
\right)\nonumber\\[2pt]
&&P''_i=\frac{2\AA_1}{c^2}\left(  K'_i H' +\varepsilon_{ijk}J'_j
P'_k  - \frac 3 2 P'_i \right) .
\label{ff}
\eea
The constant $\AA_1$ is determined similarly to the NH case.

\noindent
{\bf Proposition 3.}  {\em For a given irreducible representation of the Poincar\'e  algebra with second-order Poincar\'e Casimir ${\cal C}'_1\ne 0$, the generators defined by   (\ref{ff})
close the $(3+1)$D (anti)de Sitter  algebras whenever  the
constant  $\AA_1$ is related to the value ${\cal C}'_1$ by} 
\be
\AA_1^2 = \frac {\k_1 c^2}{4 {\cal C}'_1 } =\pm\frac {c^2}{4
\tau^2 {\cal C}'_1 }.
\label{fg}
\ee

Notice that within our approach light-like representations of the Poincar\'e algebra cannot provide deformations to (anti)de Sitter representations. 
 
\noindent
{\em Proof:}  As in the previous section  the isotopy subalgebra $\mathfrak{h}^{(1)}$ of an event, remains invariant under this deformation, thus we
only compute the   Lie brackets for the translation generators that
belong to $\mathfrak{p}^{(1)}$; these directly give (\ref{fg}):
$$
\comm{P''_i}{P''_j}=-\frac{4\AA_1^2 }{c^4}\,\varepsilon_{ijk}\,{\cal
C}'_1\,J'_k\equiv
-\frac{\k_1}{c^2}\,\varepsilon_{ijk} \,J''_k
 \qquad
\comm{H''}{P''_i}=\frac{4\AA_1^2 }{c^2}\,{\cal
C}'_1\,K'_i\equiv \k_1 K''_i  .
$$


\subsect{Deformation of Poincar\'e representations}

Further to the above algebraic results, which are in agreement with those
established in \cite{Gilmore,Rosen}, we  explore the
deformation of representations. 

By taking into account  the
deformed generators (\ref{ff}) with the solution  (\ref{fg}), it can
be proven that the (anti)de Sitter invariants (\ref{fb})  are
expressed in terms of the  Poincar\'e Casimirs (\ref{bb}) as
\be
 {\cal C}''_1=-\frac{9\k_1}{4c^2}+\k_1\,\frac{{\cal C}'_2}{{\cal
C}'_1} \qquad
 {\cal C}''_2= - \frac{\k_1}{4c^2} \,\frac{{\cal C}'_2}{{\cal
C}'_1} .
\label{ffll}
\ee
Since  this Poincar\'e deformation is only possible when  
$ {\cal C}'_1\ne 0$, the  result (\ref{ffll}) means that, in general,
any massive Poincar\'e particle  gives rise generically to an (anti)de Sitter
representation with ${\cal C}''_1\ne 0$, with a singular behaviour when ${\cal C}'_2=9\, {\cal C}'_1/(4c^2)$ which gives ${\cal C}''_1=0$. As far as the second Casimir it is clear that ${\cal C}'_2=0$ leads to ${\cal C}''_2=0$. 

In particular, let us consider     the
irreducible Poincar\'e representation  of mass $m\neq0$ and spin $s\neq0$ in the  momentum
$\>p$-space~\cite{Bacrybis}: 
\bea
&&H'=c\sqrt{\>p^2+m^2c^2}\qquad P'_i=p_i 
\qquad
J'_i=\varepsilon_{ijk}\,p_k\,\frac{\partial}{\partial p_j}+S_i 
\nonumber\\[2pt]
&&K'_i=\frac 1c \left(  \sqrt{\>p^2+m^2c^2} \,\frac{\partial}{\partial
p_i} + \frac{\varepsilon_{ijk}\,p_j 
S_k}{mc+\sqrt{\>p^2+m^2c^2}}\right) 
 \label{fk}
\eea
where $S_i$ are   spin operators such that the eigenvalue of $\>S^2$ is
$s(s+1)$; note that   $p^0=c\sqrt{\>p^2+m^2c^2}$.
This   representation is characterized by the
eigenvalues of both Casimirs ${\cal C}'_l$ (\ref{bb}):
\be 
{\cal C}'_1=-m^2c^2\qquad {\cal C}'_2= {m^2} \>S^2 .
  \label{fl}
\ee 
Then    we find that the  eigenvalues of the
  (anti)de Sitter Casimirs (\ref{ffll}) read
\be 
{\cal C}''_1=- \frac{\k_1}{c^2} \left(\frac 9 4 + \>S^2 \right)   \qquad
{\cal C}''_2=  \frac {\k_1}{4 c^4 } \,\>S^2
  \label{fn}
\ee 
so that, in this case ${\cal C}''_1\ne 0$ for any non-vanishing spin.
Finally, by taking $\AA_1 =\sqrt{-\k_1}/(2m)$,   the deformed generators
(\ref{ff}) yield to the following (anti)de Sitter representation
\bea
&&\!\!\!\!\!\!\! 
 H''=\frac {\sqrt{-\k_1}}{
mc} \,\sqrt{\>p^2+m^2c^2} \left( \>p \cdot\frac{\partial}{\partial
\>p } +\frac{3}{2} \right) \label{fm}\\[2pt]
&&\!\!\!\!\!\!\! 
  P''_i=\sqrt{-\k_1} \left( \frac {p_i}{m c^2}
\left( \>p \cdot\frac{\partial}{\partial \>p }  +\frac{3}{2} \right)+
m\,\frac{\partial}{\partial p_i}-
  \frac{\varepsilon_{ijk}\,p_j\,
S_k}{mc^2+c\sqrt{\>p^2+m^2c^2}} \right) 
\nonumber
\eea
with $\>J''=\>J'$ and $\>K''=\>K'$.


\sect{Outlook: Beyond classical deformations}

The deformation method proposed in \cite{expansion} for the  $(2+1)$D kinematical algebras 
has been generalized  in order to be applicable to higher dimensions and to any
Lie algebra, and has been explicitly applied to the 
$(3+1)$D Galilei algebra. The method works in the two physically
meaningful `deformation directions' which lead  
either to the Poincar\'e  algebra or to the
  NH ones. Deformations going from Poincar\'e to both de Sitter algebras
have   been  recovered as well. Furthermore, beyond the pure algebraic
constructions,   we have presented a preliminary study of deformations of
operators and representations  which may deserve deeper development. 

For the deformation  from Galilei to Poincar\'e, the  method as described here does not provide all
the Poincar\'e representations, but only the `light-like' ones associated to the
vanishing of both Poincar\'e Casimirs. To extend the method  so as to
allow the general representations of the Poincar\'e algebra  remains as an open problem.

Clearly, the  same procedure also holds
for the rank-two deformation going from Galilei to the  Euclidean algebra
$\euclid\equiv \mathfrak{iso}(4)$. If we replace in {\em all} the
Poincar\'e expressions given in section 4, the negative curvature
$\k_2=-1/c^2$ by a positive one (set $c$ equal to a pure imaginary complex number),   then we would
obtain all the Euclidean structures. The
deformation $U(\gal)\to \euclid$  would lead to similar deformed
generators  and conditions  as those characterized by the proposition 1.  In
this case, the symmetric space  ${\cal S}^{(1)}= ISO(4)/SO(4)$ is the 4D
flat Euclidean space and  ${\cal S}^{(2)}=ISO(4)/(\R\otimes SO(3))$ is a
positively curved and rank-two 6D space of  lines in the Euclidean space. 
Deformation of Galilean representations are characterized by the
deformed Euclidean Casimirs ${\cal C}'_l$ whose eigenvalues are again   equal
to zero. Summing up, we have obtained  the following rank-two  Galilei
deformations:

\medskip
{\footnotesize{
\hfill
\begin{tabular}{ccccc}
4D Euclidean&  &$(3+1)$D Galilei & 
&$(3+1)$D Poincar\'e\\ 
 $\euclid\equiv  \mathfrak{iso}(4)$ &$\longleftarrow$ &  $U({\gal})\equiv U(\mathfrak{iiso}(3))$&$\longrightarrow$ & $\poinc\equiv
 \mathfrak{iso}(3,1)$\\ 
$\k_1=0$, $\k_2>0$& &$\k_1=0$, $\k_2=0$ ($c=\infty$)&
&$\k_1=0$, $\k_2=-1/c^2$\\ 
${\cal C}'_1=0$, ${\cal C}'_2=0$  & &${\cal C}_1\ne 0$, ${\cal
C}_2\ne 0$ & &${\cal C}'_1=0$, ${\cal C}'_2=0$
\end{tabular}
\hfill}} 
\medskip

In the same way, similar rank-one deformations to those described in section
6   but now starting from
$U(\euclid)$ instead of $U(\poinc)$ can  also be  obtained. They   would
give rise to either $\mathfrak{so}(5)$ $(\k_1>0)$  or 
$\mathfrak{so}(4,1)$ $(\k_1<0)$, with underlying homogeneous spaces ${\cal
S}^{(l)}$   given in this order by: (i)   the   4D sphere 
${\cal S}^{(1)}= SO(5)/SO(4)$ with radius $R$ and   curvature $\k_1=1/R^2$,
and the Grassmannian ${\cal S}^{(2)}= SO(5)/(SO(2)\otimes SO(3))$ with $\k_2>0$;
(ii)   the 4D   hyperbolic space ${\cal S}^{(1)}= SO(4,1)/SO(4)$,
with   $\k_1=-1/R^2$ and the corresponding grassmannian ${\cal S}^{(2)}= SO(4,1)/(SO(1,1)\otimes SO(3))$
with   $\k_2>0$. In both processes the eigenvalue of the Euclidean Casimir
${\cal C}'_1$ must be different from zero and deformation of Euclidean
representations  are determined by the deformed Casimirs ${\cal C}''_l$
(\ref{ffll}) with $c$ purely imaginary (say
$c={\rm i}$, so that $\k_2=+1$). Hence the rank-one deformations studied   
here (covering  sections 5 and 6) are summarized as:

{\footnotesize{
\begin{tabular}{ccccc}
 \\ 4D Sphere  & &4D Euclidean  & &4D
Hyperbolic\\
$  \mathfrak{so}(5)$ &$\longleftarrow$ &$U(\euclid)\equiv
U(\mathfrak{iso}(4))$&$\longrightarrow$ &$ 
 \mathfrak{so}(4,1)$\\ 
$\k_1=+1/R^2$, $\k_2=+1$& &$\k_1=0$, $\k_2=+1$&
&$\k_1=-1/R^2$, $\k_2=+1$\\ 
${\cal C}''_1=\frac{1}{R^2}\left(\frac 94+\frac{{\cal C}'_2}{{\cal C}'_1}  
\right)$  & &${\cal C}'_1\ne 0$,
${\cal C}'_2$ arbitrary& &${\cal C}''_1=-\frac{1}{R^2}\left(\frac
94+\frac{{\cal C}'_2}{{\cal C}'_1}  
\right)$ 
 \\ 
${\cal C}''_2=\frac{1}{4R^2}\frac{{\cal C}'_2}{{\cal C}'_1}  $  & & &
&${\cal C}''_2=-\frac{1}{4R^2}\frac{{\cal C}'_2}{{\cal C}'_1}  $
 \\[0.4cm]
(3+1)D Anti-de Sitter  & &(3+1)D Poincar\'e & &(3+1)D  de Sitter\\
$ \desitter_+\equiv  \mathfrak{so}(3,2)$ &$\longleftarrow$ &$U(\poinc) \equiv
U(\mathfrak{iso}(3,1))$&$\longrightarrow$ &$  \desitter_-\equiv \mathfrak{so}(4,1)$\\
$\k_1=+1/\tau^2$, $\k_2=-1/c^2$ & &$\k_1=0$, $\k_2=-1/c^2$&
&$\k_1=-1/\tau^2$, $\k_2=-1/c^2$\\
${\cal C}''_1=-\frac{1}{\tau^2}\left(\frac 9{4c^2}-\frac{{\cal C}'_2}{{\cal
C}'_1}  
\right)$  & &${\cal C}'_1\ne 0$,
${\cal C}'_2$ arbitrary& &${\cal C}''_1=\frac{1}{\tau^2}\left(\frac
9{4c^2}-\frac{{\cal C}'_2}{{\cal C}'_1}  
\right)$ 
 \\ 
${\cal C}''_2=-\frac{1}{4\tau^2 c^2}\frac{{\cal C}'_2}{{\cal C}'_1}  $  & & &
&${\cal C}''_2=\frac{1}{4\tau^2 c^2}\frac{{\cal C}'_2}{{\cal C}'_1}  $
\\[0.4cm]
(3+1)D Oscillating NH & &(3+1)D Extended Galilei& 
&(3+1)D Expanding NH  \\
$\newhooke_+\equiv \mathfrak{t}_6(\mathfrak{so}(2)\oplus
\mathfrak{so}(3))$&$\longleftarrow$& 
$U(\overline{\gal})\equiv 
U(\overline{\mathfrak{iiso}(3)})$&$\longrightarrow$&$
\newhooke_-\equiv \mathfrak{t}_6(\mathfrak{so}(1,1)\oplus
\mathfrak{so}(3))$\\ 
$\k_1=+1/\tau^2$, $\k_2=0$ & &
$\k_1=0$, $\k_2=0$  & &$\k_1=-1/\tau^2$, $\k_2=0$ \\
${\cal C}'_1=0$, ${\cal C}'_2=0$  & &$m\ne 0$  & &${\cal C}'_1=0$, ${\cal
C}'_2=0$\\[0.4cm]
\end{tabular}
}} 
\medskip

It is worth remarking that
each of the above deformations introduces a new parameter $\k$ in the
initial Lie algebra and associated structures which, from the very beginning of our
construction, plays the role of a {\em constant curvature} of some underlying
symmetrical homogeneous space. In particular, in the deformation from the Galilean
spacetime to the relativistic one, this process can alternatively be seen as the
introduction of a {\em fundamental scale} $c$  ($\k_2=-1/c^2$). The sequence of these
classical deformations has a natural end when the semisimple Lie algebras
$\mathfrak{so}(p,q)$ are reached.   In this respect, we would like to stress that
the aforementioned geometrical and physical ideas behind these constructions can
also be found
  in the framework of quantum algebras or $q$-deformations~\cite{CP}  $U_z(\mathfrak{g})$ of
a Lie algebra $\mathfrak{g}$. These also make use of its universal enveloping algebra
$U(\mathfrak{g})$ together with a    deformation parameter $z$ ($q={\rm e}^z)$ and the
classical limit $z\to 0$ leads to $U(\mathfrak{g})$. By one hand, quantum
deformations of the Poincar\'e algebra have been used to obtain the so called
`doubly special relativities' (see  \cite{luki, alu} and references
therein) for which the deformation parameter $z$ is interpreted as a second {\em
fundamental scale}   related to the Planck length; thus this process is
`similar' to the introduction of  $c$ through a Galilean deformation but
obviously goes beyond Lie structures. On the other hand, quantum algebras have
recently been shown to provide spaces of generically {\em variable   curvature}
which is governed by $z$~\cite{plb}. Hence when classical and quantum
deformations are considered in a global picture one finds a natural `common'
deformation setting which indicates the relevance of the use of universal
enveloping algebras.  


\section*{Acknowledgments}

\noindent
Comments and suggestions from a referee and from the adjudicator have been very helpful  and are gratefully acknowledged. 
{This work was partially supported  by the Ministerio de Educaci\'on y
Ciencia   (Spain, Projects  FIS2004-07913 and MTM2005-09183) and  by the Junta de Castilla y Le\'on   (Spain, Project VA013C05)}.


\end{document}